\documentclass[aps,pra,twocolumn,superscriptaddress,floatfix]{revtex4}
\usepackage{bm} \usepackage{graphicx}
\usepackage{epsfig}

\newcommand{\bra}[1]{\langle#1|}
\newcommand{\ket}[1]{|#1\rangle}

\begin{document}


\title{Dipolar Bose-Einstein
condensates with dipole dependent scattering length}



\author{Shai Ronen}
\affiliation{JILA and Department of Physics, University of Colorado, Boulder, CO 80309-0440}
\author{Daniele C. E. Bortolotti}
\affiliation{JILA and Department of Physics, University of Colorado, Boulder, CO 80309-0440}
\affiliation{LENS and Dipartimento di Fisica, Universit\'{a} di Firenze, and INFM, Sesto Fiorentino, Italy}
\author{D. Blume}
\affiliation{Department of Physics and Astronomy, Washington State University, Pullman, 
Washington 99164-2814}
\author{John L. Bohn}
\affiliation{JILA, NIST and Department of Physics, University of
Colorado, Boulder, CO 80309-0440 }
\email{bohn@murphy.colorado.edu}

\affiliation{}


\date{\today}

\begin{abstract}

We consider a Bose-Einstein condensate of polar molecules in a
harmonic trap, where the effective dipole may be tuned by an external
field. We demonstrate that taking into account the dependence of the
scattering length on the dipole moment is essential to reproducing the
correct energies and for predicting the stability of the condensate.
We do this by comparing Gross-Pitaevskii calculations with diffusion
Monte Carlo calculations. We find very good agreement between the
results obtained by these two approaches once the dipole dependence of
the scattering length is taken into account. We also examine the
behavior of the condensate in non-isotropic traps.

\end{abstract}


\maketitle


\section{Introduction \label{intro}}
Degenerate atomic quantum gases are typically very dilute
systems. Nevertheless, inter-atomic interactions strongly determine
many of the observed phenomena and their underlying physics
\cite{BEC2003,Dalfovo99}. Until recently, only short-range and
isotropic interactions have been considered. However, recent
developments in the manipulation of cold atoms and molecules have been
paving the way towards the analyses of polar gases in which
dipole-dipole inter-particle interactions are
important\cite{Doyle04,Sage05,Stan04,Inouye04,Simoni03,Meerakker05}. A
major break through has been very recently achieved by the
experimental group in Stuttgart \cite{Stuhler05}, where a
Bose-Einstein condensate (BEC) of strongly magnetic $^{52}$Cr atoms
has been realized. This experiment observed the effects of
dipole-dipole interactions on the shape of the condensate. New
exciting phenomena are expected to occur in these quantum gases since
the particles interact via dipole-dipole interactions which are
long-ranged and anisotropic. Recent theoretical analyses have shown
that the stability and excitations of dipolar gases are crucially
determined by the trap geometry
\cite{Goral00,Yi00,Yi01,Goral02,Baranov02,ODell04,Santos00,
Santos03,Nho05}.

Yi and You \cite{Yi00,Yi01} first introduced a pseudopotential
appropriate for describing slowly moving particles interacting via
short range repulsive forces and long range dipolar forces. The
dipolar long range part of their pseudopotential is identical to
the long range part of the original potential. The pseudopotential
also includes a contact (delta function) potential whose coefficient
is proportional to the scattering length. For non-polar particles, the
scattering length is solely due to short range interactions. A crucial
point that Yi and You have shown is that the scattering length is also
dependent on the long-range dipolar interaction. They have argued for
the need to take this into account when calculating condensate
properties through the Gross-Pitaevaskii equation (GPE). However, they
stopped short at actually explicitly showing how condensate properties
are influenced by the dependence of the scattering length on the
dipole moment. Rather, they have analyzed the dependence of the
condensate properties on the \textit{ratio} of the dipole moment to
the scattering length. This is appropriate if one considers a fixed
dipole moment (and scattering length) but varies the number of
particles in the trap. On the other hand, the scenario we wish to
consider is that where dipolar interactions are tuned by an external
field, in which case the dipole moment and scattering length do not
scale equally. Also, Ref.~\cite{Yi01} did not consider negative
dipole-dependent scattering lengths.

In this work we focus on a trapped gas of dipolar particles, where the
inter-particle interaction is dominated by the dipole-dipole force. A
possible realization includes an (electrically polarized) gas of
heteronuclear molecules with a large permanent electric dipole
moment. The effective dipolar interaction may be tuned by the
competition between an orienting electric field and the quantum or
thermal rotation of the molecule. For example, the $^{2}\Pi_{3/2}$
ground state of the OH molecule is completely polarized in a field of
about $10^4$ V/cm. For smaller fields, the field strength determines
the degree of polarization and the size of the dipole moment. Another
approach for tuning the dipolar interaction, by using a rotating
external field, was proposed in Ref.~\cite{Giovanazzi02}.

We consider the case of a bare (zero dipole moment) positive
scattering length, and show that as the dipole moment is increased,
this scattering length becomes smaller and, above a certain dipole
strength, negative, and then again positive after crossing a
resonance. Taking this variation of the scattering length into
account is necessary to describe the creation of new two-body bound
states. This has consequences for the basic theory, and for the
predicted stability of BECs with anisotropic interactions, especially
for polar molecules, which can have large dipole moments. To verify
that the dipole-dependent scattering length is essential to
reproducing the correct energetics, and to ascertain the validity of
this approach, we have solved the many-body Schr\"odinger equation
using the diffusion Monte Carlo (DMC) method, as well as solved the
Schr\"{o}dinger equation by direct diagonalization for two
particles. In these calculations we used a potential consisting of
short range hard wall potential and long range dipolar interaction. We
then compare these results with solutions of the GPE, in which we
employ the appropriate pseudopotential with the dipole-dependent
scattering length.

Section~\ref{form} describes the system under study while
Sec.~\ref{manybody} presents our results for dipolar gases under
spherically symmetric and cylindrically symmetric confinement.
Selected results for systems under spherically symmetric traps have
been presented in a previous paper~\cite{Bortolotti06}. Here, we
provide numerical details (see the Appendix), justify the formalism
employed in detail, and significantly extend the discussion of our
findings. Section~\ref{conclusion} concludes.

\section{Formulation \label{form}}
For $N$ identical bosons in an external trap potential
$V_{ext}(\bm{r})$ with pair-wise interaction $V(\bm{r}-\bm{r'})$, at
zero temperature, the condensate is typically described using
mean-field theory. All the particles in the condensate then have the
same wave function $\psi(\bm{r})$. As a first attempt, one may write a
Hartree-Fock equation for this wavefunction~\cite{Esry97}:
\begin{widetext}
\begin{eqnarray}
\mu\psi(\bm{r})=\left[-\frac{\hbar^{2}}{2m}\nabla^{2}+V_{ext}(\bm{r})+(N-1)\int d\bm{r'}
V(\bm{r}-\bm{r'})|\psi(\bm{r'})|^{2}\right]\psi(\bm{r}),
\label{eq:HF}
\end{eqnarray}
\end{widetext}
where $\mu$ denotes the chemical potential, $\bm{r}$ the displacement
from the trap center, and $m$ the mass of a particle. In the
following we restrict ourselves to cylindrically symmetric harmonic
traps with $V_{ext}(\bm{r})=\frac{1}{2} m (\omega_{\rho}^{2}
\rho^{2}+\omega_{z}^{2}z^{2})$.

We consider an interaction potential $V(\bm{r})$ which consists of a
dipolar interaction and a short range hard wall with cutoff radius
$b$:
\begin{eqnarray}
V(\bm{r})=\left\{ \begin{array}{ll}
d^{2}\frac{1-3\cos^{2}\theta}{r^3} & \text{if $r \geq b$} \\
\infty & \text{if $r<b$} \end{array} \right. ,  
\label{eq:pot}
\end{eqnarray}
where $d$ denotes the dipole moment (in Gaussian units), $r$ the
distance vector between the dipoles, and $\theta$ the angle between
the vector $\bm{r}$ and the dipole axis, which we take to be aligned
along the $\hat{z}$-axis of the trap. The hardwall cutoff corresponds
to hard particles with diameter $b$ (two particles cannot penetrate
each other when the distance between their centers is equal to their
diameter). Unfortunately, the short range part of the interaction
potential $V$ causes the integral in Eq.~(\ref{eq:HF}) to diverge.
Fortunately, this divergence can be cured since the condensate is very
dilute and ultra-cold, which implies that the particles in it are
moving very slowly. Thus the potential $V$ may be replaced by an
effective potential (pseudopotential) $V_{eff}$ with a milder small
$r$ behavior, which reproduces the two-body scattering wavefunction
asymptotically in the zero energy limit, and which does not lead to
divergencies when used in the Hartree-Fock mean-field description.

For a short range potential $V$, the low energy scattering amplitude
is completely determined by one parameter, the scattering length $a$,
and an appropriate pseudopotential is
$V_{eff}(\bm{r})=\frac{4\pi\hbar^{2}a}{m}\delta(\bm{r})$. In general,
the pseudopotential is chosen such that its first-order Born
scattering amplitude reproduces the complete scattering amplitude of
the original potential $V$ in the zero-energy limit. For a potential
with long range dipolar part, Yi and You \cite{Yi00, Yi01} proposed
the following pseudopotential:
\begin{eqnarray}
V_{eff}(\bm{r})=\frac{4\pi\hbar^{2}a(d)}{m}\delta(\bm{r}) + d^2\frac{1-3\cos^{2}\theta}{r^3}
,
\label{eq:pseudo}
\end{eqnarray}
where the scattering length $a(d)$ depends on the dipole moment $d$.
Note that the long range part of the pseudopotential is identical to
the long range part of the original potential. By construction, the
scattering amplitude of $V_{eff}$ calculated in the first-order Born
approximation agrees with the full zero-energy scattering amplitude of
$V$. To verify the validity of the pseudopotential for systems of
experimental interest, we compute the low energy scattering amplitude
$f(\bm{k}, \bm{k}')$ for two OH-like molecules interacting through the
model potential given in Eq.~(\ref{eq:pot}), and compare it with the
scattering amplitude computed in the first Born approximation.

\begin{figure}
\resizebox{3.4in}{!}{\includegraphics{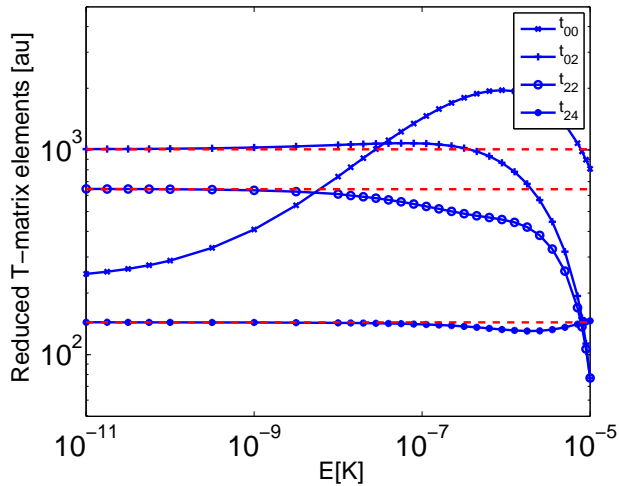}}
\caption{Absolute values of selected reduced T-matrix elements
$t_{ll'}\equiv t_{l0}^{l'0}$ (symbols) for the OH-OH model potential
as a function of the relative scattering energy, compared with the
first-order Born approximation (dashed lines). Note, the first-order
Born approximation to $t_{00}$ diverges and is not shown.
\label{scat1}}
\end{figure}

The low energy scattering amplitude in the presence of the long-range
dipolar interaction may be expanded in partial waves:
\begin{eqnarray}
f(\bm{k},\bm{k'})=4 \pi \sum_{l m, l' m'}t_{lm}^{l'm'}(k)Y^{*}_{lm}(\hat{\bm{k}})Y_{l'm'}(\hat{\bm{k'}}),
\label{Tmatrix}
\end{eqnarray}
with $t_{lm}^{l'm'}(k)$ the reduced T-matrix elements. The anisotropic
dipolar potential implies that $f$ depends on the incident and
scattered directions $\hat{\bm{k}}$ and $\hat{\bm{k'}}$. The reduced
T-matrix elements are related to the usual T-matrix elements
$T_{lm}^{l'm'}(k)=\bra{lm}\bm{T}(k)\ket{l'm'}$ by
$t_{lm}^{l'm'}(k)=\frac{ T_{lm}^{l'm'}(k) }{2k}$. For $k\rightarrow 0$
they are energy independent, and act as generalized scattering
lengths. In particular, the scattering length $a(d)$ is given by
$t_{00}^{00}(0)$ and depends on the dipole moment $d$.

Figure~\ref{scat1} compares some reduced T-matrix elements (symbols),
computed through numerical close-coupling calculations, with the first
Born approximation \cite{Yi00,Yi01,Avdeenkov05} (dashed lines), for
the potential given in Eq.~(\ref{eq:pot}) with parameters chosen to
describe the scattering between two rigid OH-like molecules. In
particular, we have taken $m=17.00$ amu, $d=0.66$ a.u., and $b=105$
a.u. (this is a reasonable value for a molecular scattering
length). For the $t_{00}^{00}$ channel, the Born approximation to the
long range dipolar part of the potential gives no contribution, while
that to the hard-core part diverges (and is therefore not shown in the
figure). For the other channels, there is a remarkably good agreement
in the $E\rightarrow 0$ limit between the exact reduced zero-energy
T-matrix elements and those calculated in the first Born
approximation. The agreement becomes less good at finite but small
scattering energies (of the order $10^{-7}$ K). Also, $t_{00}(k)$ is
not constant (as it would be in the threshold limit) even for very low
energies of the order of $10^{-10}$ K. This suggests that, e.g., an
effective range correction \cite{Braaten01,Fu03} may become important
at finite $E$, but here we consider only the $E=0$ limit. The
$t_{00}^{00}(k=0)$ matrix element determines $a(d)$, the
dipole-dependent scattering length of the pseudopotential,
Eq.~(\ref{eq:pseudo}). In this way, the Born approximation to the
pseudopotential gives the correct $t_{00}^{00}(k=0)$ value. The fact
that $a(d)$ depends on $d$ is important, since the strength of the
dipolar interactions may be controlled by an external field. Our
analysis confirms that the pseudopotential approximation provides a
good description in regions away from resonance~\cite{Yi00} (see also
below).
 
Replacing $V$ in Eq.~(\ref{eq:HF}) with $V_{eff}$ of
Eq.~(\ref{eq:pseudo}), with the scattering length $a(d)$ determined
through numerical coupled-channel calculations for the model potential
$V$, we obtain the (time-independent) Gross-Pitaevskii equation (GPE):
\begin{widetext}
\begin{eqnarray}
\mu\psi(\bm{r})=\left[-\frac{\hbar^{2}}{2m}\nabla^{2}+V_{ext}(\bm{r})+(N-1)\int d\bm{r'}
V_{eff}(\bm{r}-\bm{r'})|\psi(\bm{r'})|^{2}\right]\psi(\bm{r}).
\label{gpe}
\end{eqnarray}
\end{widetext}

\begin{figure}
\resizebox{3.4in}{!}{\includegraphics{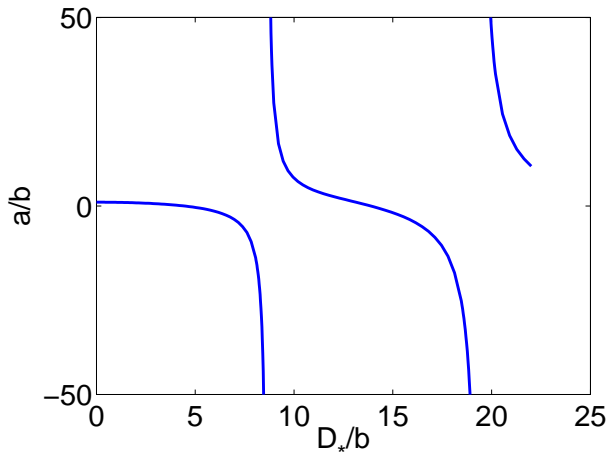}}
\caption{Scattering length $a(d)$ versus dipole length $D_*$ for the
dipolar potential with hard wall cutoff $b$ given in
Eq.~(\protect\ref{eq:pot}).
\label{scat2}}
\end{figure}

For the following, we define a dipole length $D_{*}=m d^2/
\hbar^{2}$. This is the distance at which the dipolar potential energy
equals the kinetic energy (estimated from the uncertainty relation) of
two interacting dipoles. In Fig.~\ref{scat2} we plot the ratio $a/b$
of the scattering length $a$ to the hardwall cutoff $b$ as function of
$D_{*}/b$. This provides a universal curve for the model potential
given in Eq.~(\ref{eq:pot}), which determines the scattering length
for any given cutoff $b$ and dipole length $D_{*}$. Note the
appearance of resonances, corresponding to the appearance of new bound
states. In the neighborhood of a resonance, the scattering length
tends towards $-\infty$ before and $+\infty$ after the
resonance. These resonances have been identified before in dipolar
scattering \cite{Yi00,Ticknor05}. They would occur at fields of order
MV/cm in atoms and kV/cm in heteronuclear molecules, or perhaps at
$10^5$ V/cm in atoms, if assisted by Feshbach resonances
\cite{Krems06}.

It is instructive to connect Fig.~\ref{scat2} back to concrete dipoles
that can be handled in the laboratory. Consider, for example, atomic
chromium, which is generating a lot of interest now that it has been
Bose condensed. $^{52}$Cr has a magnetic dipole moment of $6 \mu_B$,
and a sextet scattering length of 112 a.u. Identifying the hardwall
cutoff with this scattering length, chromium would appear on
Fig.~\ref{scat2} at the value $D_*/b = 0.4$, and the scattering length
would be renormalized by $\sim 0.6\%$ of its value in the absence of a
dipole moment. Indeed, this correction is already included in
two-body modeling of the Cr-Cr interaction
\cite{Werner05,Pavlovic05}. For chromium, resonances of the type shown
in Fig.~\ref{scat2} play no role.

For a heteronuclear polar molecule, however, the situation can be
quite different. Consider, for example, the OH radical, which is also
the focus of intense experimental efforts \cite{Bochinski04,Sage05}.
This molecule has a permanent electric dipole moment of 0.66 atomic
units, and therefore a huge dipole length, $D_* = 1.35 \times 10^4
a.u.$. If we assume a small cutoff, such as $b=105$ a.u., then $D_*/b
= 128$, which is way off-scale in Fig.~\ref{scat2}. In other words,
this potential supports a large number of OH-OH dimer states. These
dimer states could play a large role in the physics of a gas
composed of polar molecules.

\section{Many body and two-body calculations \label{manybody}}
The mean field GP approach to solving the many-body dynamics is
necessarily approximate. However, the full $N$-body Schr\"{o}dinger
equation with the actual interaction potential $V$,
Eq.~(\ref{eq:pot}), can be solved numerically for relatively small
$N$. For $N=2$, the Schr\"{o}dinger equation can be solved by direct
diagonalization. For $N>2$, direct diagonalization becomes
impractical and we solve the Schr\"odinger equation instead by the DMC
method~\cite{hamm94}. This section compares the results obtained by
the three methods. Details of how these schemes are implemented are
presented in the Appendix.

We note that the system is scalable: if the hardwall cutoff $b$ and
the dipole length $D_{*}$ are each scaled by a factor $K$, then the
scattering length scales by the same amount. If the harmonic
oscillator lengths $\sqrt{ \frac{\hbar}{m\omega_{z}} }$ and
$\sqrt{\frac{\hbar}{m\omega_{\rho}} }$ are also scaled accordingly,
then the entire spectrum (both the exact spectrum obtained by solving
the $N$-body Schr\"odinger equation and the spectrum obtained within
the mean field approximation) remains the same apart from scaling by a
factor $1/K^{2}$.

In what follows, unless stated otherwise, we work with an isotropic
harmonic trap: $\omega_{\rho}=\omega_{z}
\equiv\omega$. Correspondingly, the natural unit of length is the
harmonic oscillator length $a_{ho}=\sqrt{\frac{\hbar}{m\omega}}$. In
the following, we consider the two-body potential $V$,
Eq.~(\ref{eq:pot}), for two different values of $b$, i.e., $b=0.0137$
a.u. and $b=0.0433$ a.u., and varying $d$. For concreteness, OH
molecules in a trap with a frequency $\omega=2\pi$ kHz would have
$a_{ho}=5800$ a.u. so that $b=0.0137 a_{ho}$ corresponds to a
hardwall cutoff of 79 a.u.

\subsection{Energies and collapse in an isotropic trap \label{sec:energies}}
In this paper, our primary interest is in BEC's of polar molecules.
In such systems, the dipole moment is something that is directly under
the experimentalists control: in zero electric field the dipole moment
vanishes, while at higher fields its value can be continuously tuned.
For this reason, we consider properties of dipolar condensates as a
function of the dipole moment, which is taken as a substitute for the
dependence on the strength of an external electric field. One
quantity can be converted into the other, of course, via the
polarizability of the molecule.

\begin{figure}
\resizebox{3.4in}{!}{\includegraphics{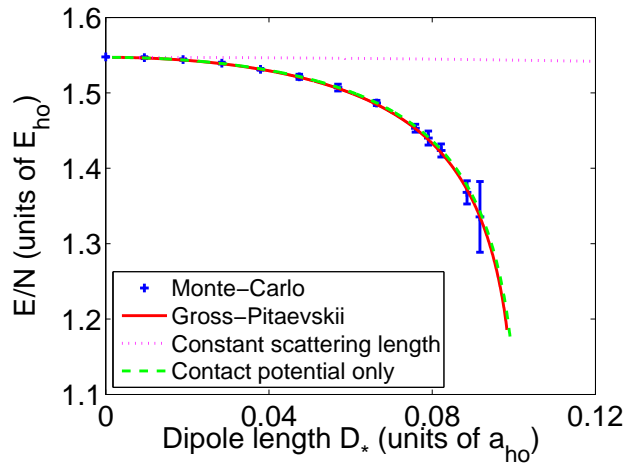}}
\caption{ Energy per particle $E/N$ as a function of the dipole length
$D_{*}$ for $N=10$. The symbols with error bars show DMC results. The
solid line is the solution of the GPE. The dotted line
shows a GP calculation without taking the dipole dependence of the
scattering length into account, setting it equal to the hardwall
cutoff. The dashed line shows a GP calculation with the dipole
dependent scattering length, but omitting the long range dipolar
term. The lines terminate at the collapse point of the condensate.
\label{wide}}
\end{figure}

Figure~\ref{wide} presents the ground state energy per particle in
units of the harmonic oscillator energy $E_{ho} = \hbar \omega$ versus
the dipole length $D_*$ in units of the harmonic oscillator length
$a_{ho}$ for $N=10$ molecules. In this figure, the solid line shows
the GP energies obtained using $V_{eff}$ with the dipole-dependent
s-wave scattering length $a(d)$. For comparison, the symbols show the
results from the DMC simulations, which solve the $N$-body
Schr\"odinger equation for the model potential $V$, Eq.~(\ref{eq:pot})
(statistical uncertainties are indicated by vertical errorbars). The
agreement is excellent, attesting to the validity of parametrizing the
GP equation in terms of the dipole-dependent scattering length
$a(d)$.

For zero dipole moment, the scattering length is equal to the hardwall
cutoff $b$. Thus the energy per particle is larger than its
noninteracting value of $1.5 \hbar \omega$. As the dipole moment
increases, the energy per particle drops. Qualitatively, this has
been explained previously by an elongation of the dipolar gas in the
$z$-direction. This allows more dipoles to encounter one another in
an attractive ``head-to-tail'' orientation. As shown below, we indeed
observe an elongation of the condensate.

However, by far the greatest influence of the dipolar interaction on
the condensate \textit{energy} comes from the dipole dependent
scattering length. To illustrate this, the dashed line in
Fig.~\ref{wide} shows the result of a GP calculation, which includes
the contact (delta function) term of $V_{eff}$ with the dipole
dependent scattering length, but omits the non-isotropic dipolar long
range term. It is quite surprising how good this approximation is, in
which the condensate is completely isotropic. For comparison, a
dotted line shows the results from a GP calculation which keeps the
long range interaction, but does not take the dependence of the
scattering length on the dipole moment into account. Instead, we use
the zero-field scattering length, i.e., $a=b$. It is clear that this
approximation greatly overestimates the energy. It is not even a very
good approximation for small dipole moments (the second derivative at
$d=0$ does not match the curvature for the ``correct'' GP solution
(solid line)).

As the dipole moment increases, the condensate quickly moves toward a
complete collapse, with $E/N \rightarrow -\infty$. This occurs because
the scattering length attains a large, negative value. It is well
known that a condensate with short range interaction only collapses
when a critical combination of particle number and scattering length
is reached, $(N-1)a=-0.57497 a_{ho}$ \cite{Dodd96,Bradley97}. This is
exactly the point where the dashed line in Fig.~\ref{wide}
terminates. It is also very nearly the point where the solid line,
which provides the most complete description of dipolar gases at the
GP level, terminates. The collapse discussed here is very different
from that usually discussed in the context of dipolar gases, which
takes the scattering length to be constant. In the latter case,
collapse can also occur due to the presence of a large dipole moment,
but (as seen from the dotted line in Fig.~\ref{wide}) this collapse
occurs for much larger dipole moments than the value predicted here by
taking the dependence of the scattering length on the dipole moment
into account. We therefore predict a kind of electric-field-induced
condensate collapse, which shares many similarities with the collapse
encountered in alkali atom BEC's near magnetic field Feshbach
resonances \cite{Bradley97,Donley01}. In particular, ramping the field
across a resonance would likely result in the creation of dimer states
of the molecules.

\begin{figure}
\resizebox{3.4in}{!}{\includegraphics{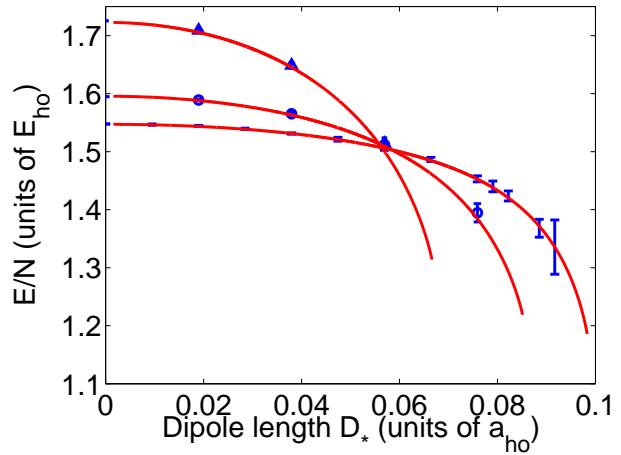}}
\caption{ Energy per particle $E/N$ as a function of the dipole length
$D_{*}$ for $N=10,20$ and $50$ (from top to bottom). Solid lines show
mean field GP energies. Dots ($N=10$), circles ($N=20$) and triangles
($N=50$) show DMC energies. Error bars indicate the statistical
uncertainties of the DMC energies.
\label{wides}}
\end{figure}

Figure~\ref{wides} shows the dependence of the energy per particle on
the dipole moment for various particle numbers, $N=10,20$ and $50$.
In all cases the conclusions are the same. Namely, the GP energies
obtained using $V_{eff}$ with the renormalized scattering length
(solid lines) are an excellent approximation to the DMC energies
(symbols). Also, the collapse occurs for smaller dipole moment as the
particle number increases, in accord with the collapse criterion.

In the present discussion we have tuned the dipole moment starting
from $d=0$. Because of the resonance structure shown in
Fig.~\ref{scat2}, it is possible to start with a large dipole moment
and a small dipole-dependent scattering length. In this case we expect
that the anisotropic long-range part of the effective potential will
carry greater weight (i.e., will have a larger impact on the
energetics and collapse). We have not considered this case in the
present work.

\begin{figure}
\resizebox{3.4in}{!}{\includegraphics{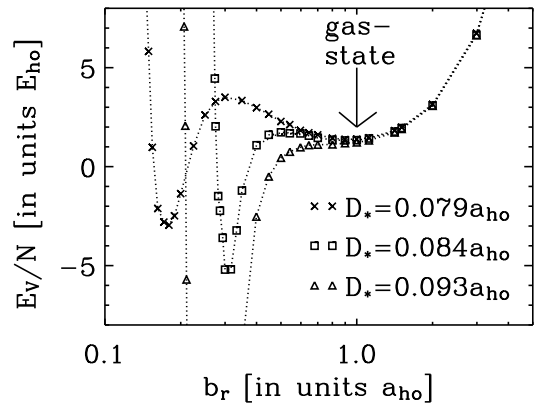}}
\caption{ Variational many-body energies $E_V/N$ for $N=20$, $b=0.0137
a_{ho}$ and $D_* = 0.079 a_{ho}$ (crosses), $D_* = 0.084 a_{ho}$
(squares) and $D_* = 0.093 a_{ho}$ (triangles) as a function of the
Gaussian width $b_r$, where $b_r=b_{\rho}=b_z$.}
\label{fig:VMC}
\end{figure}

We now investigate the stability of dipolar Bose gases by considering
the variational energy $E_V$ obtained by a variational many-body
ansatz (see the Appendix for details and notation). Symbols in
Fig.~\ref{fig:VMC} show $E_V/N$ for $b=0.0137a_{ho}$, $N=20$ and three
different values of $D_*$, i.e., $D_*=0.079 a_{ho}$ (crosses),
$D_*=0.084 a_{ho}$ (squares) and $D_*=0.093 a_{ho}$ (triangles), as a
function of the Gaussian width $b_r$, which is treated as a
variational parameter (we set $b_{\rho}=b_z=b_r$ in this stability
study). The variational parameters of $F$ (Eq.~(\ref{eq_twobody}))
are optimized for $D_*=0.079 a_{ho}$ and $0.084 a_{ho}$ by minimizing
the energy $E_V$ for fixed $b_{\rho}$ and $b_z$, i.e., for
$b_{\rho}=b_z=1$. For $D_*=0.093 a_{ho}$, we use the same values for
$p_1$ through $p_5$ as for $D_*=0.084 a_{ho}$ since no local minimum
exists for $D_*=0.093 a_{ho}$ (see below).

Figure~\ref{fig:VMC} indicates that the variational energy for
$D_*=0.079 a_{ho}$ shows a local minimum at $b_r \approx 1 a_{ho}$ and
a global minimum at $b_r \approx 0.09 a_{ho}$. The ``barrier'' at
$b_r \approx 0.3 a_{ho}$ separates the large $b_r$ region in
configuration space, where the metastable condensate state exists,
from the small $b_r$ region where bound many-body states
exist~\footnote{Many-body bound states exist for sufficiently large
$D_*/a_{ho}$ even though the two-body potential used in our DMC
calculations does not support any bound states.}. At very small
$b_r$, the variational energy becomes large and positive due to the
hardcore repulsion of the two-body potential. Figure~\ref{fig:VMC}
indicates that the energy barrier decreases with increasing $D_*$.
The dipolar gas collapses at the $D_*/a_{ho}$ value for which the
energy barrier vanishes. Our DMC calculations show that the
condensate prior to collapse is only slightly elongated, which
justifies that our stability analysis parametrizes the one-body term
$\phi$ (see Eq.~(\ref{eq_onebody})) in terms of a single Gaussian
width $b_r$ and which is consistent with our finding that the collapse
is induced primarily by the negative value of $a(d)$. We find similar
results for other $N$ values.

We emphasize that the presence of the energy barrier is crucial for our
DMC calculations to converge to the metastable condensate state for
sufficiently large $D_*/a_{ho}$ and not to the cluster-like ground
state (see also the Appendix).

\subsection{Condensate sizes and shapes\label{sizes}}
\begin{figure}
\resizebox{3.4in}{!}{\includegraphics{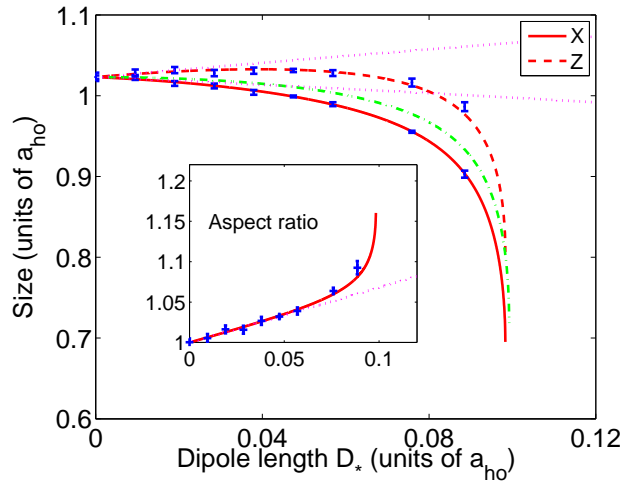}}
\caption{ Condensate sizes $X=\sqrt{<x^2>}$ and $Z=\sqrt{<z^2>}$, for
10 particles with hard wall cutoff $b=0.0137 a_{ho}$ in a spherical
trap. Solid and dashed lines show $X$ and $Z$ obtained by solving the
GP equation using $V_{eff}$ with the dipole-normalized scattering
length. Symbols with error bars show the corresponding DMC results.
The dotted lines show results of a GP calculation with constant
scattering length $a=b$. The dash-dotted line shows results of a GP
calculation with dipole dependent scattering length that omits the
long range dipolar interaction. Inset: The aspect ratio $Z/X$ for the
solution to the GP equation using $V_{eff}$ (solid line), for the
solution to the GP equation with constant scattering length $a=b$
(dotted line), and for the DMC solution to the $N$-body Schr\"odinger
equation (symbols with error bars).
\label{aspect}}
\end{figure}
The nature of the collapse is seen in more detail by looking at the
size and shape of the condensate. To this end Fig.~\ref{aspect} shows
the root-mean-square widths $Z$ and $X$ of the condensate in the $z$
and $x$ directions, for $N=10$ and $b=0.0137 a_{ho}$ (the same
parameters as in Fig.~\ref{wide}). The solid and dashed lines are
computed using the GP equation with the renormalized scattering
length, and the symbols with error bars show, as before, the results
of the DMC calculations; again, the agreement is excellent. For zero
dipole moment, the condensate is isotropic and slightly larger than
the harmonic oscillator length. As the dipole moment is turned on,
the condensate contracts slightly in the $x$ direction, and expands in
the $z$ direction. This illustrates the elongation that has been
predicted for a dipolar condensate \cite{Yi01}.

If in the effective potential we retain only the dipole-dependent
scattering length, but omit the long range dipolar term, we obtain the
dash-dotted line. This approximation describes an isotropic condensate
($Z=X$) whose size is in-between $Z$ and $X$ obtained by the more
complete description (dashed and solid lines). On the other hand, if
we retain the dipolar long range term, but set the scattering length
to a constant (its value for zero dipole moment, $a=b$), we obtain the
dotted lines. This formulation overestimates the size of the
condensate, but provides quite a good approximation to the aspect
ratio $Z/X$ (inset in Fig.~\ref{aspect}) - at least for small dipole
moments As the dipole moment is increased, the condensate approaches
the point of collapse, and the aspect ratio calculated from the
solution to the full GP equation (solid line in the inset) and the DMC
calculations (symbols in the inset) increases rapidly. The
approximation of a constant scattering length does not describe this
behavior correctly. Interestingly, relatively large anisotropy is
predicted just prior to collapse, even though the collapse mechanism
is primarily s-wave dominated.

\subsection{Energetics for dipoles in non-isotropic traps}
\begin{figure}
\resizebox{3.4in}{!}{\includegraphics{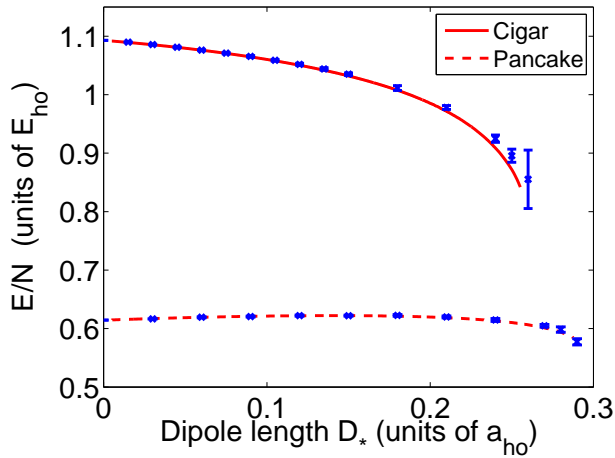}}
\caption{
Energy per particle $E/N$ obtained by solving the GP
equation using $V_{eff}$
as a function of the
dipole length $D_{*}$ for
$N=10$ in a pancake-shaped trap with $\omega_{z}/ \omega_{\rho}=10$ 
(dashed line) and in a 
cigar-shaped trap with
$\omega_{z}/ \omega_{\rho}=0.1$ (solid line). 
Symbols with error bars show the correponding DMC results.
$a_{ho}$ is defined by the shortest
dimension in each case (see text). The hard wall cutoff is $b=0.0137 a_{ho}$.
\label{pancake}}
\end{figure}
We have also explored the behavior of dipolar BECs in a pancake-shaped
trap with $\omega_{z}/\omega_{\rho}=10$ and in a cigar-shaped trap
with $\omega_{z}/\omega_{\rho}=1/10$. For the pancake trap, we define
$a_{ho}=\sqrt{\frac{\hbar}{m\omega_{z}}}$, while for the cigar trap,
$a_{ho}=\sqrt{\frac{\hbar}{m\omega_{\rho}}}$. I.e, compared to the
spherical trap of the previous sections, we elongate the trap in
either the axial or transverse directions. Figure~\ref{pancake}
indicates very good agreement between the energies calculated from the
GP equation (solid and dashed lines) and from the $N$-body
Schr\"odinger equation (symbols). In the cigar-shaped trap, a
non-vanishing dipole moment leads to a decrease of the energy compared
to the energy obtained for $d=0$. For the pancake-shaped trap, in
contrast, the energy per particle increases for small dipole moments,
and then decreases for larger dipole moments. In both traps, collapse
occurs at the $D_*$ value at which the lines in Fig.~\ref{pancake}
terminate.

We note in particular the collapse of the condensate in the pancake
trap. In \cite{Santos00} it was proposed that in a trap of such aspect
ratio, if the scattering length is zero, there should never be
collapse even for an arbitrarily large dipole moment. In our case, the
``bare'' (zero dipole moment) scattering length is positive, and so,
neglecting the scattering length dependence on the dipole moment, the
BEC should be even more resistant to collapse. Later work
\cite{Santos03}, which analyzed dipolar BECs in a trap in the limit of
$\omega_{z}/\omega_{\rho}=\infty$, suggests that a collapse should
still occur even in highly-elongated pancake traps, due to a
roton-maxon instability involving excitations with large transverse
momenta. As seen from Fig.~\ref{pancake} we do observe a collapse of
the condensate in the pancake trap, which, however, is due to the
dependence of the scattering length on the dipole moment. This
collapse mechanism is similar to that discussed above for spherical
traps (see Fig.~\ref{wide}).

\subsection{Excited states and condensate ``revival''}
The s-wave-dominated collapse shown in Fig.~\ref{wide} does not
necessarily signal the end of the story. After the scattering length
has dropped to $- \infty$, it then takes on a large positive value
(see Fig.~\ref{scat2}). In this region, the condensate is, neglecting
three-body recombination effects, perfectly stable, at least until the
scattering length again becomes large and negative. As the resonance
is crossed, a new two-body bound state appears in the potential
Eq.~(\ref{eq:pot}). Indeed, this is the first bound state of this
potential. In the following, we compare the energetics obtained by
solving the GP equation and the Schr\"odinger equation for $N=2$
particles (see Appendix) as the dipole moment is tuned across a series
of two-body resonances. Extension of this analysis to more than $N=2$
particles is complicated by the existence of the two-body bound states
of $V$, Eq.~(\ref{eq:pot}). These two-body bound states imply that
solving the $N$-body Schr\"odinger equation by the DMC method requires
the use of fixed-node techniques~\cite{reyn82}. Such a treatment is
beyond the scope of this paper.

\begin{figure}
\resizebox{3.4in}{!}{\includegraphics{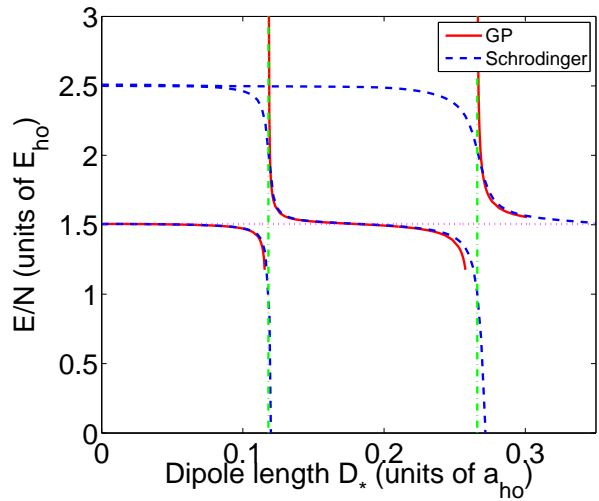}}
\caption{ $E/N$ for two particles with hard wall cutoff $b=0.0137
a_{ho}$ in a spherical trap as a function of the dipole length $D_*$
obtained by solving the GP equation (solid lines) and the
Schr\"{o}dinger equation (dashed lines). The GP line terminates at
points of collapse. For the Schr\"{o}dinger equation, the three lowest
eigenvalues are displayed (negative energies are not shown). The
dotted line shows the results from a GP calculation with a constant
scattering length $a$=$b$. The vertical dash-dotted lines indicate the
resonance positions of $V$, Eq.~(\ref{eq:pot}) (see also
Fig.~\ref{scat2}).
\label{body2}}
\end{figure}

Figure~\ref{body2} shows $E/N$ for two particles under spherically
symmetric confinement with hardwall cutoff $b=0.0137 a_{ho}$. The
exact two-body solution to the Schr\"odinger equation (dashed lines)
shows a repeated cycle of collapse and ``revival''. For comparison,
solid lines show the solution to the GP equation. In the regions away
from resonance, Fig.~\ref{body2} indicates good agreement between the
GP energies and one of the ``branches'' of the exact two-body
spectrum. To obtain this agreement, it is essential to use the
dipole-normalized scattering length, which encapsulates the
information about the formation of bound states of the system, in the
effective potential that enters the GP equation. Otherwise, if a
constant scattering length (corresponding to the zero dipole moment
case with $a$=$b$) is used in the GP formalism, the unrealistic dotted
line in Fig.~\ref{body2} is obtained. This line can only be regarded
as a good approximation in certain regions between resonances when the
dipole dependent scattering length is small. As mentioned above, for
computational reasons we have only performed this comparison for
$N=2$, but it is strongly suggestive that the same should hold for any
number of particles. By tuning the dipole moment, it should thus be
possible to observe a collapse followed by a revival of a metastable
dipolar BEC with increasing dipole strength. The results of a GP
calculation for a stability diagram of a dipolar BEC as a function of
the number of particles and the dipole moment is shown in Fig.~3 of
Ref.~\cite{Bortolotti06}.

In Fig.~\ref{body2} the resonance positions from Fig.~\ref{scat2} for
$b=0.0137 a_{ho}$ are indicated by vertical dash-dotted lines. The
solution to the two-body Schr\"odinger equation persists to slightly
larger dipole moments than the resonance position (this is most
apparent for the second resonance), because of the added kinetic
energy in the trap. By contrast, the GP solution collapses at a
smaller dipole moment, owing to the approximate collapse criterion
$(N-1)a(d)=-0.57497 a_{ho}$.

At this point we note that our solutions to the $N$-body Schr\"odinger
equation have been restricted to relatively small number of particles
($N \le 50$). It is sometimes asserted (see, e.g.,
Ref.~\cite{BEC2003}) that the GPE is valid only in the limit of large
number of particles. This restriction comes from the derivation of the
GPE from the exact Heisenberg representation equation, by replacing
the Heisenberg field operator $\bm{\hat{\Psi}}(\bm{r})$ by its
classical value $\Psi_{0}(\bm{r})$. This breaks conservation of
particle number, and can only be justified for a large number of
particles. From this point of view, one might be surprised that the
mean-field equation describes, as demonstrated here, systems with
small number of particles with high accuracy. However, it also
possible to derive the GPE from a variational ansatz which explicitly
conserves particle number \cite{Leggett01,Esry97}. This provides the
justification for applying the GPE even for a small number of
particles. In this case, it is important to keep the correct $N-1$
factor in Eq.~(\ref{gpe}), rather than, as is sometimes done, replace
it by $N$, which can only be justified as an approximation valid for a
large number of particles.

\subsection{Limits of the mean field approximation}
\begin{figure}
\resizebox{3.4in}{!}{\includegraphics{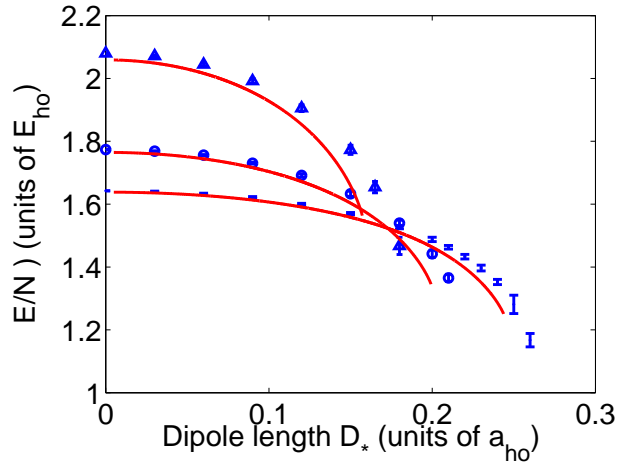}}
\caption{ Energy per particle $E/N$ versus dipole length $D_{*}$ with
hardcore radius $b=0.0433 a_{ho}$. Solid lines show the GP energies
for $N=10, 20$ and 50 particles (from top to bottom). Symbols with
error bars show the DMC energies for $N=10$ (dots), $N=20$ (circles)
and $N=50$ (triangles).
\label{N1050}}
\end{figure}
When the density of the condensate increases, one expects the mean
field GP theory to become increasingly less accurate due to
correlations among the particles. In the zero-dipole theory of short
range interactions the GP theory is the lowest order approximation in
powers of the small parameter $na^3$, where $n$ denotes the peak
density and $a$ the scattering length. The expansion parameter
becomes larger as $N$ and $a$ increase. Treating larger number of
particles in the DMC simulations increases the computational cost.
Instead, to check the applicability of the GP theory to describe dense
dipolar gases, we consider a larger particle diameter $b$.
Figure~\ref{N1050} shows a comparison between the DMC energies
(symbols with error bars) and the GP energies (solid lines) for
$b=0.0433 a_{ho}$ which is a factor $3.2$ larger than we considered in
the previous sections. A good qualitative agreement still exists, but
deviations are apparent. The GP calculations underestimate the
many-body DMC energy, and the deviations grow with increasing number
of particles $N$. The GP theory also predicts a collapse at asomewhat
smaller $d$ than the true value.

The deviations, which exist even for zero dipole moment, are
attributed by us mainly to the relatively large hard core diameter
$b=0.0433 a_{ho}$ of our ``molecules'' compared to the size of the
trap. For zero dipole moment, such effects were observed before
\cite{Blume01}. It was found that the deviations were reduced by
roughly an order of magnitude by using a modified GPE that accounts
for quantum fluctuations \cite{Braaten97,Timmermans97}. Even better
agreement was achieved when effective range corrections were taken
into account \cite{Fu03}. We expect that similar corrections would
also apply to anisotropic dipolar gases. However, Ref.~\cite{Blume01}
described condensates with short-range repulsive interactions $(a>0)$,
for which the modified GPE included a correction term calculated in
the local density approximation. In our case there is a long-range
dipolar interaction which is partly attractive, and furthermore, the
dipole dependent scattering length becomes negative for sufficiently
large $d$. As a result, the local density approximation cannot be
applied to our problem over the whole range, since a homogeneous
condensate with attractive interactions is unstable.

Finally, we note that additional beyond mean-field corrections are
expected to become important when the dipole length $D_*$ becomes
larger. Investigation of this regime is beyond the scope of this
paper.

\section{Conclusions \label{conclusion}}
In conclusion, we have considered a Bose gas with tunable dipolar
interactions, controlled by an external field, and have shown that
taking into account the dependence of the scattering length on the
dipole moment is essential for the correct description of the system
within the mean field GP method, using the pseudo-potential of Yi and
You \cite{Yi00}. By comparison with DMC simulations we have
highlighted both the accuracy of the GP method for dipolar gases with
low densities, and the existence of deviations for dipolar gases with
higher densities. The mean field theory was shown to predict
correctly the condensate size. It also describes well condensates in
cylindrical and pancake traps. It was shown that as the dipole moment
is increased, the scattering length decreases and becomes negative as
one approaches a resonance corresponding to the formation of a
two-body bound state. This effect is the major contribution for the
eventual collapse of the condensate at a critical dipole moment. This
is particularly significant for a pancake trap, in which case, failing
to take the dependence of the scattering length on the dipole moment
into account, collapse would occur only due to a roton-maxon
instability. As the dipole moment is increased further, past the
two-body resonance, the scattering length $a(d)$ becomes positive and
we see a ``revival'' of a stable condensate.

\begin{acknowledgments}
SR gratefully acknowledges financial support from an anonymous fund,
and from the United States-Israel Educational Foundation (Fulbright
Program), DCEB and JLB from the DOE and the Keck Foundation, and DB
from the NSF under grant No. PHY-0331529.
\end{acknowledgments}

\appendix*
\section{}
In this appendix we discuss the numerical techniques used in the three
different methods of this paper: solution to the Gross-Pitaevskii
equation, solution to the $N$-body Schr\"odinger equation by the DMC
method, and solution to the two-body Schr\"odinger equation.

\subsection{Solution of the Gross-Pitaevskii equation}
Equation~(\ref{gpe}) with the pseudopotential Eq.~(\ref{eq:pseudo}) is
an integro-differential equation. The integral term $d^2\int
d\bm{r'}\frac{1-3\cos^{2}\theta}{|\bm{r}-\bm{r'}|^{3}}
|\psi(\bm{r'})|^{2}$ needs special attention due to the apparent
divergence of the dipolar pseudopotential at small distances. The
integral does converge if one performs the angular integration
first. For small distances, the density $|\psi(\bm{r'})|^{2}$ may be
considered linear around $\bm{r'}=\bm{r}$ , and integrating the
angular part with $Y_{20} \propto 1-3\cos^{2}\theta $ gives zero. This
can also be seen by expanding the density in a multipole expansion
around $\bm{r'}=\bm{r}$. Only the $Y_{20}$ term will contribute to the
integral, and for a smooth density the coefficient of this term goes,
for $|\bm{r}-\bm{r'}|\rightarrow 0$, as $|\bm{r}-\bm{r'}|^2$. Thus,
the integral converges even without a cutoff.

Following Ref.~\cite{Goral02}, the calculation of the integral can be
simplified by means of the convolution theorem:
\begin{eqnarray}
\int d\bm{r'}V_{D}(\bm{r}-\bm{r'})|\psi(\bm{r'})|^{2}= \nonumber \\
\mathcal{F}^{-1}\left\{\mathcal{F}[V](\bm{k})\mathcal{F}[|\psi|^{2}](\bm{k})\right\},
\label{integral}
\end{eqnarray}
where $\mathcal{F}$ is the Fourier transform. The Fourier transform of
the dipolar potential may be performed by expanding $\exp(i\bm{k}
\cdot \bm{r})$ in a series of spherical harmonics and spherical Bessel
functions (the usual expansion of a free planar wave in free spherical
waves), where only the $Y_{20}$ term gives a non-zero
contribution. The result is:
\begin{eqnarray}
\mathcal{F}[V](\bm{k})=\frac{4\pi}{3}(3 \cos^{2}\alpha-1),
\end{eqnarray}
where $\alpha$ is the angle between the momentum $\bm{k}$ and the
dipole direction. The Fourier transform of Eq.~(\ref{integral}),
$\mathcal{F}(|\psi|^{2})$ is numerically evaluated by means of a
standard fast Fourier transform (FFT) algorithm, and multiplied by
$\mathcal{F}[V](\bm{k})$. Computation of the kinetic energy is also
accomplished to spectral accuracy through a FFT of the wave-function
and multiplication by $k^2$ in momentum space, followed by an inverse
FFT.

The ground state of the system was obtained by the usual wave-function
propagation in imaginary time. The propagation was 
implemented using a 4'th order adaptive step Runge-Kutta method. 

This procedure is very computationally intensive. We have recently
developed an improved method, which we also partly used in this work,
and which we describe elsewhere.

\subsection{Diffusion Monte Carlo simulations}
The many-body Hamiltonian $H$ for a dipolar gas under external
confinement reads
\begin{eqnarray}
H=\sum_{j=1}^N \left[ -\frac{\hbar^2}{2m} \nabla_j^2 
+ \frac{1}{2} m \left( \omega_{\rho}^2 \rho_j^2 + 
 \omega_{z}^2 z_j^2 \right) \right]
+ \nonumber \\
 \sum_{j < k} V(r_{jk},\theta_{jk}),
\end{eqnarray}
where $V$ denotes the interaction potential given by
Eq.~(\ref{eq:pot}), $\bm{r}_j=(x_j,y_j,z_j)$ the position vector of
the $j$th dipole, $\bm{r}_{jk}$ the distance vector, i.e.,
$\bm{r}_{jk}= \bm{r}_j - \bm{r}_k $, and $\theta_{jk}$ the angle
between the vector $\bm{r}_{jk}$ and the $z$-axis. To solve the
corresponding time-independent Schr\"odinger equation $H \psi = E
\psi$ we employ the variational Monte Carlo (VMC) and the Diffusion
Monte Carlo (DMC) techniques~\cite{hamm94}. The former results in a variational
bound on the energy whereas the latter results in essentially exact
many-body energies.

To determine variational many-body energies, we write the variational
wave function $\psi_V$ as a product of one-body terms $\phi$ and
two-body terms $F$,
\begin{eqnarray}
\psi_V(\bm{r}_1,\cdots,\bm{r}_N) = 
\prod_{j=1}^N \phi(\rho_j, z_j)
\times
\prod_{j<k}^N F(r_{jk},\theta_{jk}),
\end{eqnarray}
where $\phi$ contains the Gaussian widths $b_{\rho}$ and $b_z$,
\begin{eqnarray}
\label{eq_onebody}
\phi(\rho,z)= 
\exp \left[ -\frac{1}{2} \left( \frac{\rho}{b_{\rho}}\right) ^2 - 
\frac{1}{2} \left( \frac{z}{b_z} \right) ^2 \right],
\end{eqnarray}
and $F$ the hardcore radius $b$ and the ``shape'' parameters $p_1$
through $p_5$,
\begin{eqnarray}
\label{eq_twobody}
F(r,\theta)& = & \left[1-\frac{b}{r}\right]\times\nonumber \\
& & \left[1+\frac{1}{r^{p_3}}
(p_1+p_2\cos^2\theta)+\frac{p_4}{r^{p_5}}\cos^4\theta\right].
\end{eqnarray}
To first order the parameters of $\phi$ are determined by the external
confining potential and those of $F$ by the two-body potential $V$.
For a non-interacting gas with $b=d=0$, e.g., the variational wave
function $\psi_V$ with $b_{\rho}=b_z=1$ and $p_1=p_2=p_4=0$ is an
exact solution to the many-body Schr\"odinger equation. For a finite
hardcore radius $b$, the term in the first pair of square brackets on
the right hand side of Eq.~(\ref{eq_twobody}) coincides with the
low-energy $s$-wave scattering wave function for two particles
interacting through a spherically symmetric hardcore potential with
radius $b$, i.e., this term ensures that the many-body wave function
goes to zero as the distance $r$ between two particles approaches the
hardcore radius $b$. The functional form of the angle-dependent term
of $F$ [term in the second pair of square brackets on the right hand
side of Eq.~(\ref{eq_twobody})] is motivated by the shape of the
essentially exact two-body wave functions for non-vanishing $d^2$,
which is obtained numerically using B-splines (see Appendix,
subsection \ref{subsec:Bspline}).

The variational parameters $b_{\rho}$, $b_z$ and $p_1$ through $p_5$,
collectively denoted by $\bm{p}$, are optimized for each $N$,
$\omega_{\rho}$, $\omega_z$ and $D_*$ by minimizing the energy $E_V$,
\begin{eqnarray}
E_V(\vec{p}) = 
\frac{\langle \psi_V(\bm{p})|H|\psi_V(\bm{p}) \rangle}
{\langle \psi_V(\bm{p})| \psi_V(\bm{p}) \rangle},
\end{eqnarray}
where the integration over the $3N$ coordinates $\bm{r}_1, \cdots,
\bm{r}_N$ is performed using Metropolis sampling. Our
parametrization of the variational wave function $\psi_V$ provides an
excellent description of weakly-interacting gases with small $n b^3$
and $n D_*^3$, where $n$ denotes the condensate density at the trap
center. As $b/a_{ho}$ or $D_*/a_{ho}$ increase, i.e., as the gas
becomes more strongly-interacting, the VMC energy $E_V$ recovers an
increasingly smaller fraction of the essentially exact DMC energy.

The optimized variational wave function $\psi_V$ is subsequently used
as a guiding function in our DMC calculations with importance
sampling. The DMC technique solves the
 many-body Schr\"odinger
equation for the ground state
 energy and for structural properties
by starting with an initial ``walker distribution'', which can be
thought of as a stochastic representation of the many-body wave
function, and by then projecting out the lowest stationary
eigenstate through propagation in imaginary time. To treat
high-dimensional systems (here with as many as 150 degrees of
freedom), the short-time Green's function
 propagator is evaluated
stochastically. The resulting DMC energies are essentially exact,
i.e., they are independent of $\psi_V$, with the only uncertainty
stemming from the finite time step used in the propagation. For the
DMC energies reported, the time step errors are smaller than the
statistical uncertainties. Usage of a good guiding function, i.e.,
our optimized $\psi_V$, is essential for the DMC algorithm to be
numerically stable. If $\psi_V$ coincides with the exact many-body
wave function--- as is the case for non-interacting inhomogeneous
gases---, the resulting DMC energies have vanishing variance and thus
vanishing errorbars. Interactions, and correspondingly approximate
$\psi_V$, introduce statistical uncertainties in the DMC energies,
which can be reduced by increasing the computational efforts. We
calculate structural expectation values using a descendant weighting
scheme~\cite{kalo74,liu74,barn91}, which in principle eliminates any
dependence of the structural expectation values on $\psi_V$. However,
our structural expectation values may be slightly biased by $\psi_V$
since there is a tradeoff between the statistical uncertainty and the
lengths of the ``side walks'' used in the descendant weighting scheme.

The standard DMC algorithm outlined above describes the ground state.
To describe metastable condensates, i.e., excited many-body states
with gas-like character, our DMC simulations take advantage of the
topology of the underlying configuration space. The metastable
condensate state, characterized by large interparticle distances, is
separated from bound many-body states, characterized by small
interparticle distances, by a ``barrier'' (see Section
\ref{sec:energies}). In this respect, the topology of the
high-dimensional configuration space at hand is similar to the
configuration space of a one-dimensional double-well potential with
large barrier. For large enough barrier, the two regions in
configuration space correspond to two effectively orthogonal Hilbert
spaces. This implies that DMC walkers with initial coordinates
corresponding to a metastable gas-like state have a vanishingly small
probability to tunnel into the region of configuration space
corresponding to molecular-like bound states and that the simulations
consequently converge to the metastable condensate state and not to
energetically lower-lying cluster states. The presence of the barrier
is thus crucial for describing metastable condensate states with
negative scattering length by the DMC algorithm.

\subsection {Numerical solution of Schr\"{o}dinger equation for
two dipoles \label{subsec:Bspline}}
For two dipoles in a trap interacting through the potential of
Eq.~(\ref{eq:pot}), the Schr\"{o}dinger equation is separable in
relative distance and center of mass coordinates. The center of mass
equation describes the harmonic motion of the two dipoles as a whole
in the trap, and has the usual harmonic oscillator ground state. The
relative distance equation has cylindrical symmetry and was solved
numerically by expanding the wavefunction on a basis set of
two-dimensional B-splines, with the appropriate boundary
conditions. The full Hamiltonian matrix in this basis was constructed
and diagonalized.

\bibliography{biblo}

\end{document}